\documentclass[twocolumn,english,showpacs,aps,pra]{revtex4-1}
\pdfoutput=1
\usepackage[latin1]{inputenc}
\usepackage[T1]{fontenc}
\usepackage{amsmath}
\usepackage{amssymb}
\usepackage{mathtools}
\usepackage{siunitx}
\usepackage{hyperref}
\hypersetup
{   pdfsubject={Theoretical Physics},
    pdftitle={Manipulation of qubits in non-orthogonal collective storage modes},
    pdfstartview=FitH,
    colorlinks=true}
\newcommand{\lref}[1]{Eq.~\eqref{#1}} % \eqref laver en reference med parenteser omkring (til brug ved ligninger.)

\newcommand{\fref}[1]{\figurename~\ref{#1}}
\makeatletter %Følgende gør, at subscripts bliver ikke-kursiv. Anvendes X_|<subscript>|. Erstattes evt. med X_{\mathrm{<subscript>}}.
\begingroup
\catcode`\_=\active
\protected\gdef_{\@ifnextchar|\subtextup\sb}
\endgroup
\def\subtextup|#1|{\sb{\textup{#1}}}
\AtBeginDocument{\catcode`\_=12 \mathcode`\_=32768 }
\makeatother

\begin{document}
\title{Manipulation of qubits in non-orthogonal collective storage modes}
\author{Jonas Refsgaard}
\author{Klaus Mølmer}
\email{moelmer@phys.au.dk}
\affiliation{Lundbeck Foundation Theoretical Center for Quantum System Research, Department of Physics and Astronomy, University of Aarhus \\ DK-8000 Aarhus C, Denmark}
\date{\today}

\begin{abstract}
\noindent We present an analysis of transfer of quantum information between the collective spin degrees of freedom of a large ensemble of two-level systems and a single central qubit. The coupling between the central qubit and the individual ensemble members may be varied and thus provides access to more than a single storage mode. Means to store and manipulate several independent qubits are derived for the case where the variation in coupling strengths does not allow addressing of orthogonal modes of the ensemble. While our procedures and analysis may apply to a number of different physical systems, for concreteness, we study the transfer of quantum states between a single electron spin and an ensemble of nuclear spins in a quantum dot.
\end{abstract}


\maketitle

\section{Introduction}

Storage of quantum states in the collective quantum degrees of freedom of a large ensemble of identical two-level systems combines the advantage of the long coherence lifetime of microscopic systems with the strong coupling to auxiliary quantum degrees of freedom due to collective enhancement. Numerous implementations have been studied, from storage of optical states of light in optically dense atomic ensembles \cite{Polzik,Julsgaard,Reim} to the coupling of single electronic spin states with nuclear spin ensembles \cite{Taylor, Giedke, Petta, Vandersypen}, and transfer of superconducting qubit states via quantized cavity fields to rotational states in molecular ensembles \cite{Rabl}, and to collective states in electronic and nuclear spin states \cite{Wesenberg,Sorensen,Kubo,Zhu}.

For applications in quantum computing and communication one needs the capability to store and manipulate several qubits. While this can be achieved by application of a separate ensemble for each qubit, there has also been a number of proposals to identify orthogonal collective excitation modes in a single ensemble systems for multimode storage and manipulation.

Multimode storage of light has thus been demonstrated in inhomogeneously broadened media \cite{Afzelius,LeGouet}, while use of different molecular \cite{Tordrup1} and atomic \cite{Brion, Pedersen1} states can be used to collectively store separate qubit or oscillator states. With spatially extended media, one has the possibility to apply ideas from holography and store excitation patterns with different spatial periodicities \cite{Tordrup2,Wesenberg} and also to use the dephasing caused by inhomogeneities to address different collective spin superposition states \cite{Hahn,Wu}. The weak coupling of nuclear spins to their surrounding host material and to external perturbations make them ideal candidates for long time storage, but the same weak coupling makes it difficult to establish and address an independent set of collective nuclear spin modes, unless, as in \cite{Wu}, one may use the phase evolution of an electron spin ensemble to develop collective superposition phases and only subsequently transfer them to the nuclear spin ensemble.

In this paper, we assume that some controllable inhomogeneity is available in the coupling strengths of our central qubit to the ensemble members, but this inhomogeneity is insufficient to switch the coupling between orthogonal and independent collective excitation degrees of freedom. We present a method that allows us to effectively define and address independent qubits in such a system. The analysis may apply for a variety of physical systems, but for concreteness we consider a single electron, captured in a generic semiconductor quantum dot, and coupled through the hyperfine magnetic dipole interaction to the nuclear spins in the bulk of the quantum dot \cite{Taylor}. We assume that the spatial wavefunction of the electron can be manipulated by external fields, and since the spin-spin coupling depends on the electron density at the site of every nucleus, we can hence manipulate the precise form of the  collective coupling to the nuclear spins.

In Sec. \ref{an_electron}, we present the physical model and we introduce the notation and concepts used in the manuscript. In Sec. \ref{multiple_oscillators}, we identify different, but non-orthogonal nuclear spin modes, that couple to the electron spin. In Sec. \ref{adressing}, we present a procedure that allows addressing of two orthogonal modes by suitably timed sequences of interaction with the non-orthogonal spin modes. In Sec. \ref{sec-fidel}, we investigate the fidelity of our protocol, and we propose a high probability heralding scheme which significantly improves the fidelity. In Sec. \ref{discussion} we conclude  and discuss our results.

\section{An electron spin in a nuclear spin bath}
\label{an_electron}
\begin{figure}[htbp]
\centering
\includegraphics[width=0.6\columnwidth]{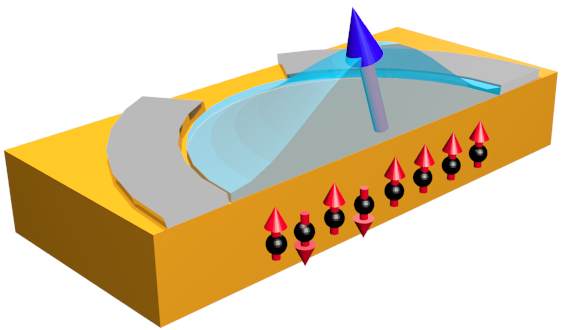}
\caption{\small Quantum dot realized in the tight potential well in the plane of a heterojunction of two semiconductor materials with different bandgaps. In a lightly doped or even intrinsic material with very few impurities, the mobility of the trapped electrons can be high and transverse confinement within the plane can be established by the potential from electrodes on the surface of the semiconductor. The figure indicates that the electron wave function, and hence, the interaction between the electron and nuclear spins are spatially dependent.}
\label{prik}
\end{figure}

We consider the situation depicted in \fref{prik}, of the spin degree of freedom associated with a spatially confined electron wavefunction. The spin of the electron interacts with external magnetic fields and with the magnetic field created by the nuclear spin ensemble within the spatial extent of the electronic state.

The magnetic dipole-interaction between the electron spin $\mathbf{S}$ and the $i$th nuclear spin $\mathbf{I}_i$ is given by
\begin{align}
H_i &= g_i \mathbf{S} \cdot \mathbf{I}_i \nonumber \\
&= g_i \biggl\{ \frac{1}{2} \bigl( S^+ I^-_i + S^- I^+_i \bigr) + S_z I_{z,i} \biggr\} ,
\end{align}
where the interaction strength, $g_i$, depends on the geometry of the quantum dot. In particular $g_i\propto \lvert \psi_g(\mathbf{r}_i)\rvert^2$, where $\psi_g(\mathbf{r}_i)$ is the spatial wave function of the electron at the position of the $i$th nucleus. We assume, that the $N$ nuclear spins are perfectly polarized in the $-z$-direction, i.e. $m_{I,i}=-I_0$. In that limit the total Hamiltonian, $H_g=\sum_i H_i$, can be written as
\begin{align}
H_g= \sqrt{\frac{N I_0}{2}}\bar{g}\bigl(S^+ b_g + S^- b^\dagger_g\bigr)+\gamma \mu_B \mathbf{S} \cdot \mathbf{B}_|OH| .
\label{ham}
\end{align}
In Eq. (\ref{ham}) we have implemented the \emph{Holstein-Primakoff-approximation}, which describes the collective nuclear spin excitation by bosonic creation and annihilation operators. This approximation assumes that the number of excitations is substantially less than the total number of spins, and as we will consider only 0,1 and 2 excitations within an ensemble of thousands of spins, it is perfectly valid. In Eq. (\ref{ham}), $\bar{g}=\sqrt{(1/N)\sum_i |g_i|^2}$, is the rms-value of the coupling strengths $\{g_i\}$, $\gamma$ is the effective Land\'{e}-factor of the electron, which depends on the quantum dot geometry and host material, and the \emph{Overhauser-field}, $\mathbf{B}_|OH|$, is given by the expression
\begin{align}
\mathbf{B}_|OH|\equiv \frac{1}{\gamma \mu_B}\sum_{i=1}^N g_i I_{z,i} \mathbf{\hat{z}} .
\end{align}

With the above definitions we can write the creation and annihilation operators explicitly as
\begin{align}
b^\dagger_g = \frac{1}{\sqrt{N}}\sum_{i=1}^N \frac{g_i^*}{\bar{g}}\sigma^+_i \quad \mathrm{and} \quad b_g = \frac{1}{\sqrt{N}}\sum_{i=1}^N \frac{g_i}{\bar{g}}\sigma^-_i .
\label{op}
\end{align}
Let us now consider the effect of these operators in detail. We have assumed perfect polarization of the nuclear spins, so we can define our ground state as $\vert 0 \rangle = \vert 0_1 0_2 \ldots  0_N\rangle \equiv \lvert (-I_0)_1 (-I_0)_2 \ldots (-I_0)_N \rangle$, while, for a state with a single excited spin, we write, $\vert\ldots 1_i\ldots \rangle \equiv \vert \ldots (-I_0+1)_i\ldots \rangle$.

If we apply the Holstein-Primakoff creation operator to the ground state we get
\begin{align}
\vert 1_g \rangle = b^\dagger_g \vert 0 \rangle = \frac{1}{\sqrt{N}}\sum_{i=1}^N \frac{g_i}{\bar{g}}\vert 0_1 0_2 \ldots 1_i \ldots 0_N \rangle,
\label{exc}
\end{align}
which can be interpreted as a single collective excitation in the nuclear spin-ensemble. This excited state is characterized by the set $\{g_i\}$ that describes the pattern of the amplitudes, with which the individual nuclear spins have been flipped.

If we inspect the Hamiltonian in Eq. (\ref{ham}), we see that it consists of two different terms. The first term is a flip-flop-interaction that conserves the total spin projection. With the Holstein-Primakoff approximation for the nuclear spin states, this interaction is equivalent to the Jaynes-Cummings Hamiltonian, effectively coupling a two-level system (the electron spin) to a harmonic oscillator (the excitations in the nuclear spin-ensemble). The second term in the Hamiltonian describes the electron spin precessing in a magnetic field, and we will include also an external field, such that $\mathbf{B}_|OH|\rightarrow\mathbf{B}_|OH|+\mathbf{B}_|ext|=\mathbf{B}_|eff|$. By adjusting the external field the flip-flop-interaction can be turned on and off resonance.

\section{Multiple Oscillators}
\label{multiple_oscillators}
As proposed in \cite{Taylor}, the nuclear spin degree of freedom has a very long lifetime and therefore the quantum information represented by the electron spin may be transferred to the nuclear ensemble for robust long-time storage. In this manuscript our goal is to store more than a single qubit in the same nuclear spin ensemble. This is possible if we can address orthogonal collective spin wave modes, as e.g., done in \cite{Hahn,Tordrup2,Wesenberg}. Here, however, the effectively coupled nuclear spin ensemble is confined to the spatial volume occupied by a single electron, and we do not have the same means to address orthogonal plane wave modes.

\begin{figure}[htbp]
\centering
\includegraphics[scale=1]{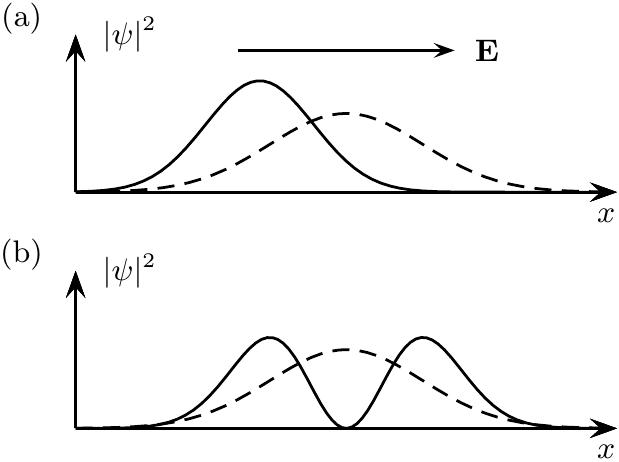}
\caption{\small This figure illustrates how the electronic wavefunction of an  electron in the ground state (dashed line) of a quantum dot can be modified (solid line) by the application of an electric field (a), and by excitation of the electron (b).}
\label{ford}
\end{figure}

We may, however, perturb the spatial wavefunction of the electron by applying an electric field with a component parallel to the plane of the quantum dot or we may excite the electron to another motional bound state in the quantum dot, see also \fref{ford}. In this altered state, the wavefunction is modified, $\psi_g({\bf r}) \rightarrow \psi_h({\bf r})$, and since the coupling strengths between the electron spin and the individual nuclear spins depend on the electron spatial probability distribution the electron spin hence couples to a different collective spin degree of freedom characterized by the coupling strengths, $\{h_i\}\propto |\psi_h({\bf r}_i)|^2$. The Hamiltonian which governs this new interaction, $H_h$, is otherwise analogous to \lref{ham}, but with the creation and annihilation operators replaced by the ones corresponding to the new oscillator.

By controlling the electron spatial wavefunction we thus have a choice between two different storage modes, but unless the modes can be addressed independently of each other, this does not in itself provide the ability to use the nuclear spin ensemble as a register for the storage of two different qubits. The requirement of independence is met if the mode operators for the two spin modes commute, and in particular,
\begin{align}
\left[b^{\phantom{\dagger}}_g , b^\dagger_h \right] =  0 .
\end{align}
Using the Holstein-Primakoff-approximation, we obtain
\begin{align}
\left[b^{\phantom{\dagger}}_g , b^\dagger_h \right] = \frac{1}{N\bar{g}\bar{h}}\sum_{i=1}^N g_i^* h^{\phantom{*}}_i,
\label{kom}
\end{align}
which in turn is equal to the overlap $\langle 1_g \vert 1_h \rangle$ between the single excitation quantum states, $\lvert 1_\alpha \rangle=b^\dagger_\alpha \lvert 0 \rangle, \alpha=g,h$.

\begin{figure}[htbp]
\centering
\includegraphics[scale=1]{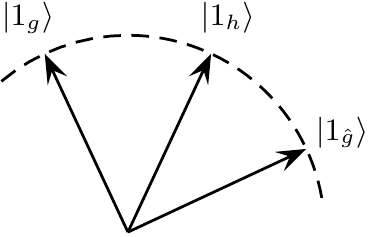}
\caption{\small A geometric illustration of the mode orthogonality problem. We can selectively address one of two different oscillators, $g$- and $h$. Their first excited states span a two-dimensional Hilbert-space but they are not orthogonal. We identify a scheme to construct and address a linear superposition of the two modes, $\hat{g}$, that is orthogonal to $g$.}
\label{cirkel}
\end{figure}

The states $\lvert 1_g \rangle$, $\lvert 1_h \rangle$ span a two dimensional Hilbert space,  and it is possible to construct a superposition of the states $\lvert 1_g \rangle$ and $\lvert 1_h \rangle$ that is orthogonal to $\lvert 1_g \rangle$, see \fref{cirkel}. We label this state $\lvert 1_{\hat{g}}\rangle$, and using the \emph{Gram-Schmidt-method} we find an explicit form of $\lvert 1_{\hat{g}}\rangle$:
\begin{align}
\lvert 1_{\hat{g}} \rangle &=(1-\langle 1_g \vert 1_h \rangle^2 )^{-1/2}( \lvert 1_h \rangle - \langle 1_g \vert 1_h \rangle \lvert 1_g \rangle) .
\end{align}

The overlap $\langle 1_g \vert 1_{\hat{g}}\rangle=0$ ensures that the operator $b^\dagger_{\hat{g}}$ creating $\lvert 1_{\hat{g}}\rangle$ from the fully polarized spin state, indeed commutes with the collective raising and lowering operators (\ref{op}) for our original spin wave mode,
\begin{align}
\left[b^{\phantom{\dagger}}_g , b^\dagger_{\hat{g}} \right]=0 ,
\end{align}
and it is possible to exchange quantum states between the electron spin and the $g$-oscillator without modifying the state of the $\hat{g}$-oscillator.

\section{Adressing the Quantum Memory}
\label{adressing}
We will use the $g$-oscillator mode as the \emph{read-in/read-out}-mode for direct transfer of qubit states between the electron spin and the nuclear spin ensemble. When a quantum state has been transferred to the $g$-oscillator, subsequent transfer of the state into the $\hat{g}$-oscillator can be accomplished with the repeated application of the pulse sequence
\begin{align}
\mathcal{U}_{\tau} \equiv e^{i H_h \tau}e^{i H_g \tau}e^{-i H_h \tau}e^{-i H_g \tau},
\label{sekvens}
\end{align}
using precisely the Hamiltonian interaction operators available to us, when the electron occupies its two possible spatial states.

The switching between the operators is obtained by changing the electronic motional state between $\psi_g$ and $\psi_h$. By adjusting the external magnetic field the flip-flop-interaction can be tuned off resonance while the electronic state is manipulated, and when the interaction is off resonance the electron spin will precess around the $z$-axis. If we let the electron precess $\pi$ radians the $S_x$- and $S_y$-operators, and therefore also the $S^+$- and $S^-$-operators, will reverse sign. When we return on resonance, we can thus also obtain the change of sign of the interaction Hamiltonian (\ref{ham}) as needed in every second application of $H_g$ and $H_h$ in (\ref{sekvens}).

If we let $\tau$ be a small time interval, we can use the \emph{Campbell-Baker-Hausdorff-formula} to rewrite the operator and, ignoring terms of order higher than $\tau^2$, we get
\begin{align}
\mathcal{U}_{\tau} \simeq e^{-[H_h ,H_g]\tau^2} .
\label{approx}
\end{align}
If we apply the sequence $N$ times, the resulting time evolution operator is
\begin{align}
\mathcal{U}_N \equiv (\mathcal{U}_{\tau})^N = e^{-N[H_h,H_g]\tau^2} = e^{-iHT} \\
\shortintertext{where}
H=\frac{[H_h,H_g]\tau}{4i} \qquad \mathrm{and} \qquad T=4N\tau .
\label{eff}
\end{align}
This means that the evolution corresponds to that caused by the effective hamiltonian $H$ in a time interval $T$, which is just defined as the total time of the sequence.

With the resonant flip-flop-interaction the two Hamiltonians write
\begin{align}
H_g=\Omega_g \bigl( S^+ b^{\phantom{\dagger}}_g + S^- b^\dagger_g\bigr) \quad, H_h=\Omega_h \bigl( S^+ b^{\phantom{\dagger}}_h + S^- b^\dagger_h \bigr) ,
\end{align}
and we can evaluate the commutator in Eq. (\ref{eff}) and calculate $H$. Defining the annihilation operator $b_{\hat{g}}=(1-\langle 1_g \vert 1_h \rangle^2 )^{-1/2}( b_h - \langle 1_g \vert 1_h \rangle b_g)$ and likewise for the corresponding creation operator we then have
\begin{align}
H=\frac{\tau}{2i}\Omega_g\Omega_h \sqrt{1-\langle 1_g \vert 1_h\rangle^2} \left( b_{g}^\dagger b_{\hat{g}}^{\phantom{\dagger}} - b_{g}^{\phantom{\dagger}}b_{\hat{g}}^\dagger \right)S_z .
\label{ideel}
\end{align}
If the electron is in an eigenstate of the $S_z$-operator, the effective hamiltonian is exactly the beam-splitter interaction between the two oscillator modes, which for the appropriate interaction time will swap not only qubit states but, in fact, any quantum states between the $g$- and $\hat{g}$-oscillators. With the electron in the spin down eigenstate, we have for example
\begin{align}
 e^{-iHT}\lvert \downarrow 1_g 0_{\hat{g}}\rangle = \cos\theta \lvert \downarrow 1_g 0_{\hat{g}}\rangle - \sin\theta \lvert \downarrow 0_g 1_{\hat{g}}\rangle ,
\label{beam}
\end{align}
where $\theta=\frac{\tau}{4}\Omega_g\Omega_h \sqrt{1-\langle 1_g \vert 1_h\rangle^2} \, T$. If we want to swap an excitation from one oscillator to the other we just set $\theta=\pi/2$. We denote the corresponding time evolution operator $\mathcal{U}_{\frac{\pi}{2}}$.

From this it is clear that the for a given $\theta$ the time needed for the transformation grows as $T\propto (1-\langle 1_g \vert 1_h\rangle^2)^{-\frac{1}{2}}$. To get a sense of how big the overlap might be, we can consider the simple case where the potential in the dot is 2D harmonic with the ground state $\psi_g(x,y)=\psi_{00}(x,y)$ and excited state $\psi_h(x,y)=\psi_{10}(x,y)$, where the subscripts counts the number of excitations in the $x$- and $y$-direction. For this system we get an overlap $\langle 1_g \vert 1_h \rangle =3^{-\frac{1}{2}}\simeq 0.58$.

It is now possible to adress both storage-modes of our quantum memory: We first read an arbitrary qubit state into the $g$-oscillator and we then swap it to the orthogonal $\hat{g}$-oscillator with the pulse sequence described above. With the qubit safely stored we can read another qubit into the $g$-oscillator without disturbing the first one, and with the swapping mechanism at our disposal we can gain random access to any of the two qubits, and using the electron-nuclear spin interaction, we can also implement quantum gates on the two-bit register.

\section{Fidelity of quantum state transfer}
\label{sec-fidel}

In this section we investigate how well our pulse sequence $\mathcal{U}_{N}$ achieves the ideal time evolution, $\mathcal{U}_{\frac{\pi}{2}}$, taking into account the higher order terms, that were neglected in \lref{approx}. The Campbell-Baker-Hausdorff formula is exact in the limit of infinitesimal $\tau$, and the purpose of the current analysis is to assess how fast, and with how few steps, we may carry out the total operation without significant loss of fidelity.

We must hence calculate the fidelity between the desired and the actually achieved final state, $f=\lvert\langle\psi\vert\mathcal{U}_{\frac{\pi}{2}}^\dagger\mathcal{U}_N\vert\psi\rangle\rvert^2$, and subsequently average this quantity over the relevant initial states of our protocol. Our Hamiltonian interaction operators conserve the total number of excited spins, and we restrict ourselves here to the Hilbert subspace with between zero and two spin excitations spanned by
\begin{align}
&\underbrace{\overbrace{\lvert \downarrow 0_g 0_{\hat{g}} \rangle, \lvert \downarrow 1_g 0_{\hat{g}} \rangle, \lvert \downarrow 0_g 1_{\hat{g}} \rangle, \lvert \downarrow 1_g 1_{\hat{g}} \rangle}^\mathcal{S}, \lvert \downarrow 2_g 0_{\hat{g}} \rangle, \lvert \downarrow 0_g 2_{\hat{g}} \rangle,}_\mathcal{T} \nonumber \\ &\lvert \uparrow 0_g 0_{\hat{g}} \rangle, \lvert \uparrow 1_g 0_{\hat{g}} \rangle, \lvert \uparrow 0_g 1_{\hat{g}} \rangle .
\label{states}
\end{align}
We will obtain the exact unitary time evolution in this subspace numerically, and we will compare it with the desired evolution, applied on states from the the relevant input space, i.e., the space of state carrying the four possible two-bit states of the spin oscillators. The bracket with the symbol $\mathcal{S}$ in (\ref{states}) encompasses this information carrying subspace, and we will average the fidelity, $f$ over a uniform distribution of input states from $\mathcal{S}$. Since the ideal operation restricts the dynamics to the subspace $\mathcal{S}$, leakage of excitation to states outside $\mathcal{S}$ populated during the exact evolution transformation will automatically reduce the fidelity.

Defining  the projection operator, $P_\mathcal{S}$ on $\mathcal{S}$, the average fidelity can be expressed as the following integral over states in $\mathcal{S}$,
\begin{align}
F=\langle f \rangle = \int_\mathcal{S} \lvert\langle\psi\vert P_\mathcal{S}\mathcal{U}_{\frac{\pi}{2}}^\dagger P_\mathcal{S}\mathcal{U}_N P_\mathcal{S}\vert\psi\rangle\rvert^2 \, dV,
\label{formel}
\end{align}
where the projection operators are not strictly needed as the states are taken from $\mathcal{S}$, and $\mathcal{U}_{\frac{\pi}{2}}^\dagger$ keeps states within the subspace.
\begin{figure}[htbp]
\centering
\includegraphics[scale=1]{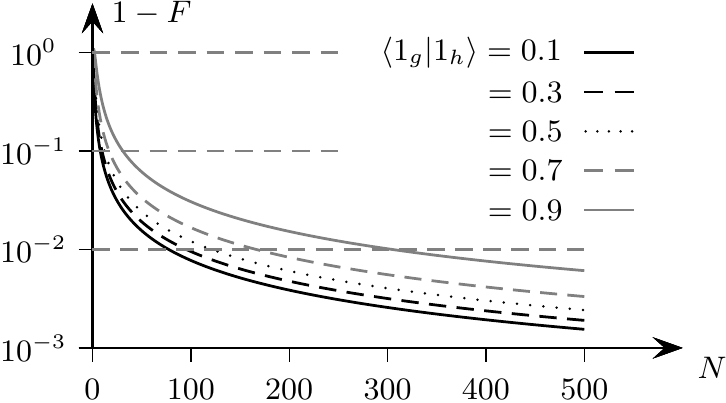}
\caption{\small The infidelity $1-F$ is plotted (on a log scale) as a function of the number of applied pulses, $N$, for five different values of the overlap $\langle 1_g \vert 1_h \rangle$.}
\label{fid1}
\end{figure}

Using the results in \cite{line1} we obtain the explicit expression
\begin{align}
F=\frac{1}{n_\mathcal{S}(n_\mathcal{S}+1)}\left\lbrace\mathrm{Tr}(MM^\dagger) + \lvert \mathrm{Tr}(M)\rvert^2\right\rbrace ,\label{formel2}
\end{align}
where $M=P_\mathcal{S}\mathcal{U}_{\frac{\pi}{2}}^\dagger P_\mathcal{S}\mathcal{U}_N P_\mathcal{S}$ is a product of projection and unitary time evolution matrices, and $n_\mathcal{S}=4$ is the dimension  of the subspace $\mathcal{S}$.

The infidelity, $1-F$, is plotted as a function of the number of repeated applications of the pulse sequence (\ref{sekvens}) in \fref{fid1}, for different values of the overlap between the two modes. As we would expect, the infidelity decays faster with $N$ when the overlap is small, which reflects the fact that it is then easier to address the two modes. In the case of a vanishing overlap, we can address the two modes independently in a direct manner, but as the overlap increases, it takes longer time and more operations, cf., the expression for $\theta$ after Eq. (\ref{beam}).

Part of the infidelity is due to population escaping the subspace $\mathcal{S}$, and ending up with state vector components along the other states listed in (\ref{states}). Since components with the electron spin up after the protocol are definitely erroneous, we suggest to perform a measurement of the electron spin after the swapping operation. With a probability that does not exceed the infidelity, plotted in \fref{fid1}, we may find the electron in the spin up state and discard the system, while if we find the electron in the spin down state, we have effectively removed the small electron spin up component from the state, and obtained a new normalized state vector belonging to the subspace $\mathcal{T}$ indicated by the bracket with the same symbol in (\ref{states}).

Thus, with a minor reduction of the success probability, we effectively enhance the desired state vector component in  $\mathcal{S}$ and thus improve the fidelity. Since the Campbell-Baker-Hausdorff formula neglects higher order terms, which contain the electron spin flip operators, we expect that the lowest order error source is associated with leakage outside of $\mathcal{T}$, and our heralding may thus significantly improve the fidelity of the protocol.

To determine the average fidelity of the heralded state transfer process, we must calculate the integral over all initial states from $\mathcal{S}$, of the squared overlap between the final (normalized) state projected into $\mathcal{T}$ and the target state, weighted by the probability of actually finding the final state in $\mathcal{T}$ \cite{line2}. The integral should finally be normalized by the average probability for the final state to be in $\mathcal{T}$:
\begin{align}
F_c &=\frac{\int_\mathcal{S} \left\lvert\langle\psi\vert P_\mathcal{S}\mathcal{U}_{\frac{\pi}{2}}^\dagger P_\mathcal{S}\frac{P_\mathcal{T}\mathcal{U}_N P_\mathcal{S} \vert\psi\rangle}{\Vert P_\mathcal{T}\mathcal{U}_N P_\mathcal{S} \vert\psi\rangle\Vert}\right\rvert^2 \Vert P_\mathcal{T}\mathcal{U}_N P_\mathcal{S} \vert\psi\rangle\Vert^2 \, dV}{\int_\mathcal{S} \Vert P_\mathcal{T}\mathcal{U}_N P_\mathcal{S} \vert\psi\rangle\Vert^2 \, dV}.
\end{align}
The numerator can be simplified, and using that $P_\mathcal{S}P_\mathcal{T}=P_\mathcal{S}$, we can express the mean conditioned fidelity as
\begin{align}
F_c =\frac{1}{n_\mathcal{S} +1}\frac{\mathrm{Tr}(MM^\dagger) + \lvert \mathrm{Tr}(M)\rvert^2}{\mathrm{Tr}(P_\mathcal{S}\mathcal{U}_N^\dagger P_\mathcal{T}\mathcal{U}_N P_\mathcal{S})},
\end{align}
where the operator $M$ is the same as in Eq. (\ref{formel2}).
\begin{figure}[htbp]
\centering
\includegraphics[scale=1]{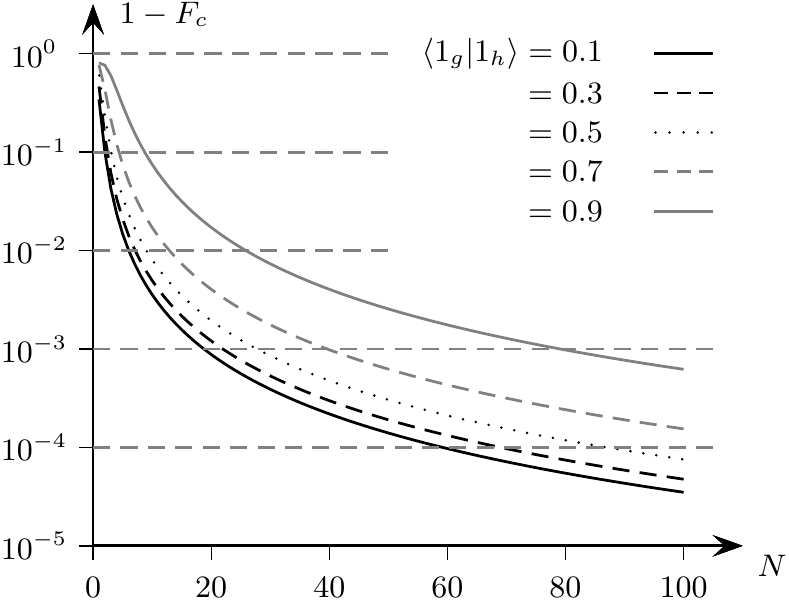}
\caption{\small The conditional infidelity $1-F_c$ is plotted as a function of the number of applied pulses, $N$, and for five different values of the overlap $\langle 1_g \vert 1_h \rangle$.}
\label{fid}
\end{figure}

We have calculated and plotted the conditional infidelity as a function of the number $N$ of pulse seqeuences in \fref{fid} for different values of the overlap $\langle 1_g \vert 1_h \rangle$.  The success probability of the heralding is higher than the unconditional fidelity, represented in \fref{fid1}, and, e.g., when  $F=\SI{99}{\percent}$, we observe a factor one hundred further reduction in the infidelity, associated with the conditioning.

We have plotted the fidelity as a function of $N$, but it is also of interest to calculate the number of pulse squences and the total time, $T$, required for the protocol at a given fidelity level. It is a straightforward exercise to show that
\begin{align}
T=\left( \frac{8\pi N}{\Omega_g \Omega_h \sqrt{1-\langle 1_g \vert 1_h \rangle^2}}\right)^{\frac{1}{2}}
\end{align}
To reach fidelity of $F_c=0.999$ for two values of the overlap, $0.1$ and $0.9$, we thus need
\begin{align}
N_{0.999}(0.1)&=19 \Rightarrow T\simeq 21.9(\Omega_g \Omega_h)^{-\frac{1}{2}} \nonumber \\
N_{0.999}(0.9)&=80 \Rightarrow T\simeq 67.9(\Omega_g \Omega_h)^{-\frac{1}{2}} .
\end{align}
As can also be read from the figure, relatively fast swapping between the oscillator modes is possible with relatively few operations. In both cases the heralding probability of finding the electron in the spin down state is close to $\SI{95}{\percent}$.

\section{Discussion}
\label{discussion}

In summary, we have shown that ensembles of spins can be used as a two-mode storage device. If two orthogonal spin excitation modes can be selectively addressed, the operation of the memory is easy, while for a coupling that addresses non-orthogonal modes, more elaborate control is needed to identify and subsequently manipulate superposition states.

We showed that for a wide range of values for the mode overlap, which may be implemented with different spatial electron wave functions in a quantum dot, it is possible to address, swap, and manipulate two qubits of information in an ensemble. With more than two choices of spatial coupling amplitudes, we imagine that the scheme may be readily generalized to more qubits. We further note, that our key theoretical component is a beam splitter operation, which swaps not only qubit states but also general oscillator states between the collective modes, and thus provides the possibility to implement controlled operation on two-qudit (multi-level) \cite{strauch,jacobs,mischuck} and on continuous variable \cite{Rudner} states. Our fidelity analysis was carried out for qubit degrees of freedom, but we expect that it is representative for the unconditioned and the conditioned fidelities attainable in qudit and continuous variable systems as well.

Our physical example dealt with the case of an electron quantum dot, where the electron spin couples to the nuclear spins in the host material within the range of the spatial electronic wavefunction. The general problem of control of non-ortogonal collective modes may occur in different microscopic and mesoscopic systems including nuclei in the vicinity of NV centers in diamond, small atomic ensembles, and superconducting elements coupled to transmission waveguides or to nanomechanical devices, and we imagine that multi-mode storage along the lines presented here may be pursued in such systems.

\end{document}